\documentstyle[aps,twocolumn,prb,epsf]{revtex}

\begin{document}

\newcommand{\be}{\begin{equation}}
\newcommand{\ee}{\end{equation}}
\newcommand{\bn}{\begin{eqnarray}}
\newcommand{\en}{\end{eqnarray}}
\def\beq{\begin{eqnarray}}
\def\eeq{\end{eqnarray}}
\def\lsim{\:\raisebox{-0.5ex}{$\stackrel{\textstyle<}{\sim}$}\:}
\def\gsim{\:\raisebox{-0.5ex}{$\stackrel{\textstyle>}{\sim}$}\:}
\def\d{\displaystyle}
\def\u{\underbar}

\draft

\twocolumn[\hsize\textwidth\columnwidth\hsize\csname @twocolumnfalse\endcsname

\title{Insulator-metal transition in the rutile-based $\rm VO_{2}$}

\author{M. S. Laad$^1$, L. Craco$^2$ and E. M\"uller-Hartmann$^2$}

\address{$^1$Department of Physics, Loughborough University, LE11 3TU, UK \\
$^2$Institut f\"ur Theoretische Physik, Universit\"at zu K\"oln,
77 Z\"ulpicher Strasse, D-50937 K\"oln, Germany}

\date{\today}
\maketitle

\widetext

\begin{abstract}
The metal-insulator transition (MIT) in paramagnetic ${\rm VO_{2}}$ 
is studied within LDA+DMFT(IPT), which merges the local 
density approximation (LDA) with dynamical mean field theory (DMFT). 
With a fixed value of the Coulomb $U=5.0~eV$, we show how the 
MIT is understood in a new picture: spectral weight transfer accompanying 
the increase in the displacement of V ion ($\perp c$) within the strong 
correlation scenario. Within this new scenario, we find good quantitative 
agreement with $(i)$ switch of the orbital occupation of $(xy,yz+zx,yz-zx)$, 
$(ii)$ thermodynamics, and $(iii)$ the one-electron spectral function in 
the metallic phase of ${\rm VO_{2}}$. We also compare our results 
for the total spectral density with other approaches which use QMC to 
solve the impurity problem of DMFT.
\end{abstract}

\pacs{PACS numbers: 71.28+d,71.30+h,72.10-d}

]

\narrowtext

\section{Introduction}

Electron correlation-driven metal-insulator transitions (MIT) in the $3d$ 
transition metal (TM) oxides remain a problem of enduring interest in
condensed matter physics.~\cite{[1]}  Vanadium oxides have proved to be
fascinating candidates in this context.~\cite{[2]} Recent X-ray absorption
(XAS) studies~\cite{[3]} on ${\rm V_{2}O_{3}}$ have led to a drastic 
revision of the traditional picture~\cite{[4]} of the MIT in this system, 
forcing a description in terms of a $S=1$, multiband model.~\cite{[5],V2O3}  
Similar behavior is also observed in pressurized 
${\rm Ca_{2}RuO_{4}}$,~\cite{[6]} suggesting that such a scenario may be 
a manifestation of the general importance of orbital correlations in TM 
oxides.

Stoichiometric vanadium dioxide (${\rm VO_{2}}$) is another example of a 
system showing a MIT with reduction of temperature. The high-$T$ metallic 
(M) phase has rutile (R) structure, which changes to monoclinic at the 
first-order MIT occuring around $T_{MI}=340$~K.  This, along with the 
observation of spin dimerization along the $c$-axis,~\cite{[7]} gave rise 
to theories linking the MIT to strong electron-lattice coupling 
(ELC),~\cite{[8]} where the important role of the antiferroelectric 
displacement of the $V$ ions ($\perp c$) was pointed out. It was suggested 
that $c$-axis dimerization of the V-V pairs would open up a gap in the 
low-$T$ insulating ($I$) state in the framework of a bonding-antibonding 
splitting. The M state, accompanied by disappearance of this dimerization, 
would result from an upward shift of the $d_{xy}$ band ($d_{\sigma}$ in 
Ref.~\onlinecite{[9]}). In a classic work, a zero-gap insulator was obtained 
within LDA;~\cite{[Allen]} this study treated structural aspects in full 
detail. However, the inability to reproduce the observed gap value suggested 
that inclusion of on-site correlations was necessary. In contrast, in a 
pioneering work, the importance of electronic correlations, along with that 
of the antiferroelectric displacement of V ions from their $c$-axis positions 
in driving the MIT was emphasized.~\cite{[9]} From a detailed perusal of 
experimental results known at that time, it was argued that $c$-axis spin 
dimerization is a consequence of the MIT driven by electronic correlations 
in concert with the antiferroelectric displacements.  Subsequently, a model 
calculation~\cite{[10]} explicitly realized this suggestion, but with
unrealistically small model parameters. On the experimental side, the 
observation of anomalous (correlation driven) features~\cite{[11],[12],[13]} 
and results from magnetic measurements~\cite{[14]} finding local moments in 
${\rm V_{1-x}Cr_{x}O_{2}}$, required a correlation based scenario. 
Simultaneously, observation of only a modest mass enhancement in the R 
phase seemed to suggest that while correlations were undoubtedly present, 
they did not play a crucial role in the MIT. In the light of this conflict, 
the origin of the PI state as well as the mechanism of the MIT is an open 
and controversial issue.

Recent (unpublished) XAS results~\cite{[15]} show that the abrupt MIT in 
${\rm VO_{2}}$ is accompanied by a hysteresis in occupation of different 
$t_{2g}$ orbitals, directly confirming the importance of orbital correlations 
in a multi-band situation.~\cite{[5]}  Recent photoemission measurements 
show a large spectral weight transfer from high- to low energies, on a scale 
of $O$(2~eV), across the MIT.  Further, this character of the 
MIT, which is driven by a sudden increase in carrier density rather than 
their mobility, is shown very clearly by time-delay measurements.~\cite{[16]}  
More recently, Lim {\it et al.}~\cite{Lim} found the MIT to be strongly first 
order by I-V measurements on thin films. Very interestingly, micro-Raman 
scattering experiment clearly shows that this electric field driven Mott
transition is {\it not} accompanied by any structural transition, in contrast
to the usual $T$ driven one, strongly suggesting the Mott-Hubbard scenario.  
These observations put particularly strong constraints on an acceptable 
theory of the MIT for ${\rm VO_{2}}$: $(i)$ it should be able to describe this 
switch in orbital occupation at the MIT, $(ii)$ the metallic phase should be 
moderately correlated (see, however,~\cite{[13]}), and, $(iii)$ the insulator 
should be of the Mott-Hubbard type, accompanied by $c$-axis dimerization of 
V$^{4+}$-V$^{4+}$ pairs, and the MIT itself should be first-order, consistent 
with Ref.~\onlinecite{[16]}.

In this article, we discuss a concrete theoretical scenario, 
extending~\cite{[9],[10]} earlier works in the light of the above 
requirements. Using the recently developed ab-initio LDA+DMFT scheme, we 
show how a consistent treatment of strong, dynamical multi-orbital 
correlations along with the antiferroelectric displacements of V in the real 
LDA bandstructure leads to a unified understanding of the MIT in 
${\rm VO_{2}}$. We also compare our results with other, recent LDA+DMFT work 
which uses quantum Monte Carlo (QMC) method to solve the impurity problem 
of DMFT. Finally, we show how very good quantitative agreement with PES, XAS, 
thermodynamic and magnetic measurements is obtained.  Our results support 
the Mott-Hubbard picture of the MIT in ${\rm VO_{2}}$, where the Peierls 
instability arises subsequent to the M-I transition.  
  
\section{Model and solution}

\begin{figure}[htb]
\epsfxsize=3.5in
\epsffile{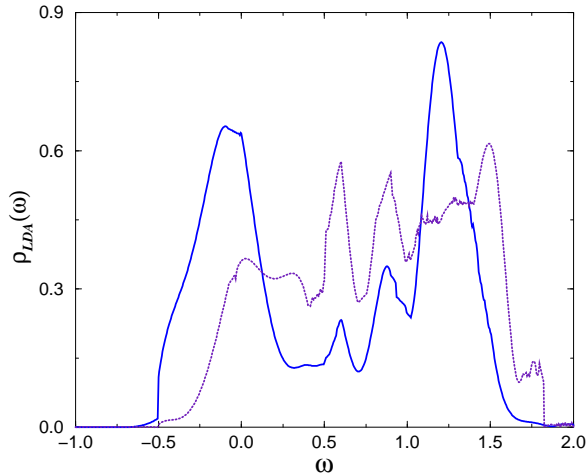}
\caption{
LDA partial density of states for the $d_{yz \pm zx}$ 
(dotted) and $d_{xy}$ (solid) orbitals, obtained from 
Ref.~\protect\onlinecite{[17]}.}
\label{fig1}
\end{figure}

We start with the LDA bandstructure of ${\rm VO_{2}}$ in the monoclinic 
crystal structure corresponding to the insulating phase.~\cite{[17]} The 
LDA DOS, computed using the known value of the antiferro-electric 
displacement of V$^{4+}$ ions ($\epsilon$, giving the splitting 
$\Delta_{\alpha\beta}$ between the $xy, yz\pm zx$ orbitals at the local level) 
(Fig.~\ref{fig1}) shows interesting features: the lowest lying $xy$ (in 
the $M$ notation) band is the most heavily populated, while the $yz\pm zx$ 
bands are less populated.  The bonding-antibonding splitting~\cite{[8]} is 
clearly visible in the results, which also show that the total bandwidth is 
about $W \simeq 2.3$~eV, contrary to much smaller previous~\cite{[8],[9],[10]} 
estimates based on model calculations. Except for their inability to 
reproduce the Mott gap, these LDA results agree with Goodenough's arguments.  
As emphasized by Mott,~\cite{[9]} it is inconceivable that $\epsilon$ alone 
could open up a charge gap ($E_{G}\simeq 0.6$~eV), this requiring proper 
incorporation of correlation effects. 

Thus, the LDA (bandstructure) part of the Hamiltonian is 

\be
H_{0}=\sum_{{\bf k} \alpha \beta\sigma}\epsilon_{\alpha\beta}({\bf k})
c_{{\bf k}\alpha\sigma}^{\dag}c_{{\bf k}\beta\sigma} + \sum_{i\alpha\sigma}\epsilon_{i\alpha\sigma}n_{i\alpha\sigma}
\ee
where $\alpha,\beta=xy,yz+zx,yz-zx$. To avoid double counting of interactions 
already treated on the average by LDA, we follow~\cite{[18]} to write

\be
H_{LDA}^{(0)}=\sum_{{\bf k}\alpha\beta\sigma}\epsilon_{\alpha\beta}
({\bf k})c_{{\bf k}\alpha\sigma}^{\dag}c_{{\bf k}\beta\sigma} 
+ \sum_{i\alpha\sigma}\epsilon_{i\alpha\sigma}^{0}n_{i\alpha\sigma},
\ee
where 
$\epsilon_{i\alpha\sigma}^{0}=\epsilon_{i\alpha\sigma}-U(n_{\alpha\bar\sigma}
-\frac{1}{2})+\frac{1}{2} J_{H}(n_{\alpha\sigma}-1)$, 
with $U, J_{H}$ as defined below, and the full Hamiltonian is,

\bn
\nonumber
H &=& H_{LDA}^{(0)}+U\sum_{i\alpha}n_{i\alpha\uparrow}n_{i\alpha\downarrow} 
+ \sum_{i\alpha\beta\sigma\sigma'} U_{\alpha\beta}
n_{i\alpha\sigma}n_{i\beta\sigma'} \\
&-& J_{H}\sum_{i\alpha\beta}{\bf S}_{i\alpha} \cdot{\bf S}_{i\beta}.
\en

Constrained LDA calculations give $U\simeq 4.5~$eV, $J_{H}=1.0$~eV and 
$U' \equiv U_{\alpha\beta}=(U-2J_{H})=2.5$~eV.  Given the uncertainty in 
the precise estimation of $U$, we choose $U=5.0$~eV in the calculations. 
Considering only the $t_{2g}$ manifold, the PM/para-orbital phase, and working
in the basis which diagonalises the LDA density matrix, we 
have $G_{\alpha\beta\sigma\sigma'}(\omega)=
\delta_{\alpha\beta}\delta_{\sigma\sigma'}G_{\alpha\sigma}(\omega)$ and 
$\Sigma_{\alpha\beta\sigma\sigma'}(\omega)=
\delta_{\alpha\beta}\delta_{\sigma\sigma'}\Sigma_{\alpha\sigma}(\omega)$.

In the $t_{2g}$ sub-basis, a DMFT solution involves $(i)$ replacing the 
lattice model by a multi-orbital, asymmetric Anderson impurity model, and 
$(ii)$ a selfconsistency condition requiring the impurity propagator to be 
equal to the local ($k$-averaged) Green function of the lattice, given by

\be
G_{\alpha}(\omega) = \frac{1}{V_{B}} \int d^{3}k 
\left[ \frac{1}{(\omega+\mu)1-H_{LDA}^{(0)}({\bf k})
-\Sigma(\omega)} \right]_{\alpha}.
\ee

Using the locality of $\Sigma_{\alpha\beta}$ in $d=\infty$, we get 
$G_{\alpha}(\omega)=G_{\alpha}^{0}(\omega-\Sigma_{\alpha}(\omega))$. Further,
since $U_{\alpha\beta}, J_{H}$ scatter electrons between the 
$xy,yz+zx,yz-zx$ bands, only the total number, 
$n_{t_{2g}}=\sum_{\alpha}n_{t_{2g},\alpha}$ is conserved in a way consistent 
with Luttinger's theorem.

To solve the impurity problem of DMFT, we use the iterated perturbation 
theory (IPT), generalized to the case of $t_{2g}$ orbitals for arbitrary 
filling.~\cite{Kot,cro2} The local propagators are given by

\be
\label{eq:G_k}
G_{\alpha}(\omega)=\frac{1}{N}\sum_{\bf k}
\frac{1}{\omega-\Sigma_{\alpha}(\omega)-\epsilon_{{\bf k}\alpha}} \;.
\ee

Local dynamical self-energies $\Sigma_{\alpha}(\omega)$ are computed 
within the extended multi-orbital iterated-perturbation theory (MO-IPT) which
generalizes the IPT for the one orbital case to the multi-orbital situation, 
for any temperature~\cite{gamnas} and band-filling.~\cite{cro2} Explicitly, 

\be
\label{eq:self}
\Sigma_{\alpha}(\omega)=\frac{\sum_\gamma A_{\alpha\gamma} 
\Sigma_{\alpha\gamma}^{(2)}(\omega)}
{1-\sum_\gamma B_{\alpha\gamma}\Sigma_{\alpha\gamma}^{(2)}(\omega)}
\ee
where, for example,

\bn
\nonumber
\Sigma_{\alpha\gamma}^{(2)}(i\omega)=
\frac{U_{\alpha\gamma}^2}{\beta^2}\sum_{n,m} G_{\alpha}^{0}(i\omega_{n})
G_{\gamma}^{0}(i\omega_{m})G_{\gamma}^{0}(i\omega_{n}+i\omega_{m}-i\omega)
\en
and
$G_{\alpha}^{0}(\omega)=
\frac{1}{\omega+\mu_{\alpha}-\Delta_{\alpha}(\omega)}$.
The multi-orbital dynamical bath $\Delta_{\alpha}(\omega)$ function 
is determined using the DMFT self-consistent condition 
which requires that the local inpurity Green function to be equal to the 
local Green function of the lattice (Eq.~(\ref{eq:G_k})). Finally, 
in Eq.~(\ref{eq:self}), 

\be
A_{\alpha\gamma}=\frac{n_{\alpha}(1-2n_{\alpha})
+D_{\alpha\gamma}[n]}{n_{\alpha}^0(1-n_{\alpha}^0)}
\ee
and,  
\be
B_{\alpha\gamma}=\frac{(1-2n_{\alpha})U_{\alpha\gamma} + \mu -\mu_{\alpha}}
{2U_{\alpha\gamma}^{2}n_{\alpha}^0(1-n_{\alpha}^0)} \;.
\ee
Here, $n_{\alpha}$ and $n_{\alpha}^0$ are particle numbers determined from 
$G_{\alpha}$ and $G_{\alpha}^{0}$ respectively, and 
$D_{\alpha\gamma}[n]=\langle n_{\alpha}n_{\gamma} \rangle$ 
is calculated using~\cite{Kot} 

\be
\label{eq:nn}
\langle n_{\alpha}n_{\gamma} \rangle =
\langle n_{\alpha}\rangle \langle n_{\gamma} \rangle 
-\frac{1}{U_{\alpha\gamma}\pi} \int f(\omega)
Im [\Sigma_{\alpha}(\omega)G_{\alpha}(\omega)]d\omega \;.
\ee
The last identity in Eq.~(\ref{eq:nn}) 
follows directly from the equations of motion for 
$G_{\alpha}(\omega)$. These equations satisfy the Luttinger sum rule, and 
reproduce the first few moments of the spectral functions~\cite{Fulde}, 
guaranteeing the correct low- and high energy behavior of the propagators. As 
shown by Kakehashi {\it et al.}~\cite{Fulde} the IPT for arbitrary filling 
can be derived in a controlled way by truncating the memory function matrix 
in a Mori-Zwanzig projection formalism. The above equations are solved 
selfconsistently respecting these constraints.

Below, we use the following strategy to study the Mott M-I transition. Since 
we do not expect much correspondence between changes in bare LDA quantities 
with those affected in non-trivial ways by dynamical electronic correlations, 
we choose the bare LDA DOS for the ``insulating'' solution (we do {\it not}
change this DOS to study the MIT) at low $T(\simeq 75$~K) in our calculation.
To access the MIT as a function of $T$, and driven by change of orbital 
occupation, we monitor the fully renormalized spectral functions, their 
renormalized occupations, along with changes in the renormalized 
antiferroelectric displacement ($\Delta_{\alpha\beta}$). To be 
precise, we start with a trial value of $\epsilon$. The multi-orbital 
correlations have two important effects in this situation. Given a trial 
value of $\epsilon$, we expect $U_{\alpha\beta}$ to renormalize its bare 
value (the effective ``field'' in the $t_{2g}$ sector) via multi-orbital 
Hartree shifts, leading $(i)$ to changes in the orbital occupations, and 
more importantly, $(ii)$ to large changes in spectral weight transfer (driven 
by {\it dynamical} nature of strong local correlations) upon small changes 
in $\Delta_{\alpha\beta}$, leading to stabilization of the second~\cite{[9]}
(metallic) DMFT solution. In other words, this large SWT drives the abrupt 
MIT around a critical value of $\Delta_{\alpha\beta}=\Delta_{c}$, as is 
indeed confirmed  below.

\section{Numerical Results}

\begin{figure}[htb]
\epsfxsize=3.4in
\epsffile{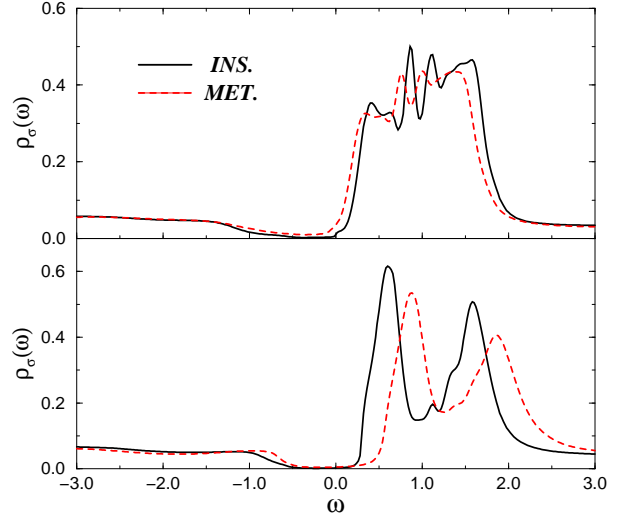}
\caption{
$t_{2g}$ partial density of states using LDA+DMFT 
for $U=5.0~eV$ in the PI (solid) and PM (dashed) phases of ${\rm VO_{2}}$. 
The upper panel corresponds to $d_{yz\pm zx}$ orbitals, and the lower panel, 
to the $d_{xy}$ orbital.} 
\label{fig2}
\end{figure}  

In Fig.~\ref{fig2} (solid lines), we show our result for the partial DOS in 
the insulating case for $T=0.0075$~eV. Notice that the charge gap, 
$E_G \simeq 0.5$~eV, consistent with the optical data.~\cite{[20]} In 
accordance with the orbital splitting assignment from LDA+U, the lower-lying 
$xy$ band is more localized in the solid. The orbital occupations in the MI 
state are computed to be $(n_{\parallel},n_{\pi}=0.36, 0.32, 0.32)$, quite
different from the XAS estimations. From 
the computed spectral functions, the renormalized value of the distortion 
(caused by $\Delta_{\alpha\beta}$) is found to be 
$\Delta_{\alpha\beta}^{I} \simeq 0.17$~eV. Decreasing the trial values of 
$n_{\parallel}$ (leading to {\it decrease} in $\Delta_{\alpha\beta}^{I}$) 
leads to a large transfer of spectral weight via DMFT in the multi-orbital 
situation, and, at a critical $n_{\parallel}$, to an abrupt (first-order)
transition to a metallic phase (Fig.~\ref{fig2}, dashed lines, computed for
$T=0.03$~eV). Interestingly, the $d_{\parallel}$ DOS still represents 
insulating behavior, while the $d_{\pi}$ DOS exhibits characteristic 
metallic behavior, providing an explicit realization of the {\it two-fluid} 
scenario~\cite{[2]} proposed earlier to describe the general problem of MIT.  
In accord with the above argument, the renormalized 
$\Delta_{\alpha\beta}^{M} \simeq -0.29$~eV, corresponding to a reduction in 
$\Delta_{\alpha\beta}$, in full agreement with Mott.~\cite{[9]} The difference 
$\delta=\Delta_{\alpha\beta}^{I}-\Delta_{\alpha\beta}^{M}=0.46$, in good 
agreement with the value of $0.5-0.6$ from cluster calculations.~\cite{[9]} 
Moreover, this agrees with the orbital splitting assignment in the M 
phase,~\cite{[8],[9],[17]} where the $d_{\pi}$ band(s) lies lower than the 
$d_{\parallel}$ band. Finally, the orbital occupations in the M phase are 
found to be $(n_{\parallel},n_{\pi})=(0.34, 0.33, 0.33)$ respectively, 
in very good semiquantitative agreement with recent XAS results. 

Based on these results, the abrupt MIT in ${\rm VO_{2}}$ is understood as 
follows. A decrease in the LDA value of $n_{\parallel}$ leads to an effective 
reduction in $\Delta^{I}$ (caused by Hartree shifts due to $U_{\alpha\beta}$)
leading to an effective {\it increase} in $n_{\pi}$, and, more importantly, 
to large changes in dynamical SWT from high- to low energy, driving the first 
order MIT. The converged values of $n_{\pi}$ (obviously different from 
the trial values) vary in a way qualitatively consistent with indications 
from recent XAS measurements.  Thus, in our picture, the MIT is driven by
$(i)$ abrupt changes in $n_{\parallel},n_{\pi}$ and $\Delta_{\alpha\beta}$ 
due to large change in dynamical SWT, and, $(ii)$ an increase in the carrier 
concentration, rather than their mobility, consistent with time-delay 
measurements.~\cite{[16]}

\begin{figure}[htb]
\epsfxsize=3.4in
\epsffile{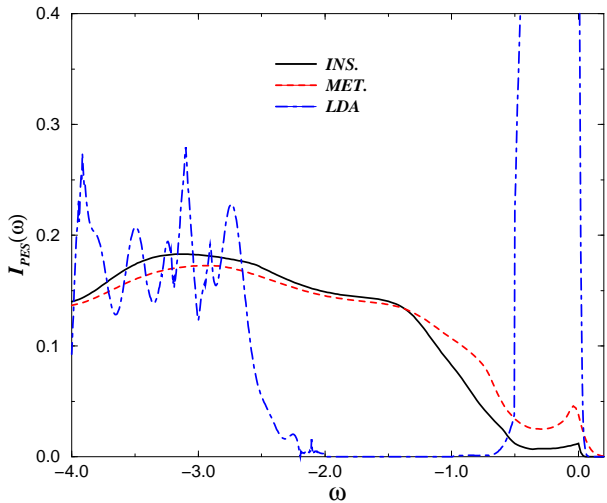}
\caption{(color online) 
The integrated photoemission lineshapes for the PI (solid) and PM 
(dashed) phases within LDA+DMFT for $U=5.0~eV$.  The LDA result is the 
dot-dashed line. } 
\label{fig3}
\end{figure}

Given the relatively wide $t_{2g}$ bands in the LDA (contrast with 
Ref.~\onlinecite{[10]}, where the bandwidths, as well as $U, U_{\alpha\beta}, 
J_{H}$ were severely underestimated), correlation induced mass enhancement 
is found to be moderate.  Indeed, from the self-energy (not shown), the 
average effective mass enhancement factor is $m^{*}/m_{0} \simeq 3.0$, in
full accord with that extracted from reflectance studies.~\cite{[21]} Clearly,
this enhanced mass comes solely from the $d_{\pi}$ band (this is the only 
band showing ``metallic'' behavior).~\cite{[9]}  Further, the high-$T$ static 
spin susceptibility is enhanced by the same factor of $3$ coming from 
renormalization effects of multi-orbital correlations, in contrast to the 
Stoner mechanism discussed before in Ref.~\onlinecite{[9]}. These results 
can be interpreted to mean a moderate renormalization caused by the 
additional screening provided by the wide $d_{\pi}$ bands in the M phase.

In Fig.~\ref{fig3}, we show the integrated photoemission (PES) spectra 
calculated in both the I and the M phases. Comparing our results with 
published experimental work,~\cite{[12]} very good semiquantitative agreement 
between theory and experiment is seen.  In this context, we emphasize that a 
certain amount of caution should be exercised while comparing theory to 
experiment because $(i)$ reduced co-ordination at the surface would normally 
enhance correlation effects, decreasing the DOS at $E_{F}$, and $(ii)$ the 
surface of ${\rm VO_{2}}$ is known to show insulating features 
(Ref.~\onlinecite{[12]}). Incorporating spin-Peierls effects in the PI 
phase will also change the PES lineshape in the PI phase somewhat; this 
is left for the future. With these caveats in mind, the overall agreement 
is strkingly good, and, in particular, the transfer of spectral weight 
across the MIT is faithfully reproduced by theory.

In particular, in our picture, the small peak in the metallic phase is 
{\it not} a quasicoherent Fermi liquid (FL) peak, but part of the tail of the 
$d_{yz\pm zx}$ bands (from Fig.~\ref{fig2}) pulled down below $E_{F}$ 
in the PM phase, as described above. More detailed orbital-selective
spectroscopic investigations are needed to confirm this 
picture.~\cite{EPL-ml} This is also consistent with the absence of a 
$T^{2}$ term in the $dc$ resistivity, which is known~\cite{[13]} to follow 
a linear-in-$T$ behavior for $330~{\rm K} < T < 840~{\rm K}$.  We notice 
that a FL peak in PES would contradict the incoherent, linear-in-$T$ 
(non-FL) resistivity.  In any case, in correlated systems on the verge 
of a Mott transition, such a FL resonance should be visible, if at all, 
only at very low temperatures, and is rapidly washed off with increasing $T$. 

\section{Comparison with LDA+DMFT(QMC) results}

\begin{figure}[htb]
\epsfxsize=3.5in
\epsffile{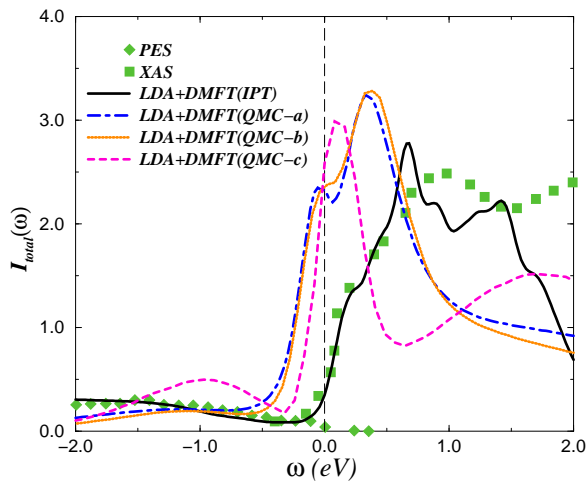}
\caption{
Comparison of theoretical LDA+DMFT(IPT) and 
LDA+DMFT(QMC) result for the total one-electron spectral function in the 
metallic phase of ${\rm VO_2}$ to the experimental results taken from 
Ref.~\protect\onlinecite{PES} (for PES) and from 
Ref.~\protect\onlinecite{XAS} (for XAS). The LDA+DMFT(QMC) results 
are taken from: Ref.~\protect\onlinecite{[AG]}~$(a)$ [$T=770$~K], 
Ref.~\protect\onlinecite{[Li-cond]}~$(b)$ [$T=770$~K], and 
Ref.~\protect\onlinecite{[Liebsch]}~$(c)$ [$T\approx 500$~K].} 
\label{fig4}
\end{figure}

We turn now to a comparison of our work with other, recent approaches which 
use QMC to solve the impurity problem of DMFT.~\cite{[AG],[Liebsch],[Li-cond]} 
First, we clearly see in Fig.~\ref{fig4} that the QMC calculations 
resolve a huge peak near $E_{F}$. Though a miniscule pseudogap can be 
discerned in Ref.~\onlinecite{[AG]}, a direct comparison between {\it both} 
LDA+DMFT(QMC) results with the corresponding LDA ones clearly shows that 
the many-body renormalised DOS, $\rho(E_{F})$ at the Fermi surface, coincides 
with the corresponding LDA DOS, even at $T=770$~K. (Note that the QMC 
calculations  done at $T \approx 500$~K~\cite{[Liebsch]} are also pinned to the
{\it same} value at $E_{F}$!).
This is clearly problematic: let us recall that the Friedel Luttinger sum 
rule implies $\rho(E_{F})=\rho_{LDA}(E_{F})$ only at $T=0$~\cite{[MH]} in 
$d=\infty$.  Further, in a DMFT framework, this pinned (at $T=0$) feature 
actually {\it ceases} to exist above the lattice coherence scale, $T_{coh}$,
which is quite small near a correlation-driven MI transition.  In the 
LDA+DMFT framework, such a pinned feature can only be associated with a 
correlated FL peak. Its persistence up to $T=770~K$ (QMC) is thus in direct 
conflict with the bad-metallic, linear-in-$T$ resistivity observed from 
much lower $T(=340~{\rm K})$ in ${\rm VO_{2}}$.  Optical measurements
further confirm this:~\cite{[20]} no coherent Drude peak is observed in 
the PM phase.

The pinning observed in LDA+DMFT(QMC) work is thus very hard to justify, 
both on theoretical and experimental grounds.  Further, explicit consideration
of inter-site V-V interactions (dimerisation) within cluster DMFT would 
normally cause a low-energy pseudogap to appear around $E_{F}$, making any 
pinning impossible, even at $T=0$.  Clearly, our results do not suffer from 
this conflict, and are moreover in excellent quantitative accord with PES/XAS 
results, as well as with the effective mass enhancement estimated from 
optics/thermodynamics, as discussed before.  To compare various approaches, 
we show in Fig.~\ref{fig4} our result, together with those obtained from 
LDA+DMFT(QMC)~\cite{[AG],[Liebsch],[Li-cond]} in comparison with recent 
experimental results.  While our result quantitatively describes the 
full spectral function from $-3.0\le\omega\le1.2$~eV, the LDA+DMFT(QMC) 
result does not capture the detailed lineshape correctly over the whole 
range.

What are the sources of the discrepancy between our results and the 
LDA+DMFT(QMC) ones?
There are two important differences between our approach compared
to the LDA+DMFT(QMC) works: The LDA+DMFT(QMC) papers derive the many-body 
density of states (DOS) in the two phases (dimerised/metallic) by using the 
corresponding LDA DOS {\it separately} for each phases, and using DMFT(QMC) 
to study dynamical correlation effects in each phase. Thus, the two phases 
can be studied separately, but it is difficult to see how the first-order 
transition between the two could be derived this way.

We argue that this is problematic because the MI transition in the actual 
correlated system is driven by large changes in transfer of dynamical 
spectral weight in response to small changes in the {\it renormalised} LDA 
parameters. In a multi-orbital system, these (the AFE distortion in 
${\rm VO_2}$) are themselves modified in a priori unknown (sometimes 
non-trivial) ways by electronic correlations. Our approach is to search 
for an instability from the Mott insulating (MI) phase (corresponding to 
one solution of the DMFT equations) to the paramagnetic metallic phase 
(the second solution of the DMFT equations) as a function of the 
renormalised AFE distortion, as explained above.
  
\begin{figure}[htb]
\epsfxsize=3.4in
\epsffile{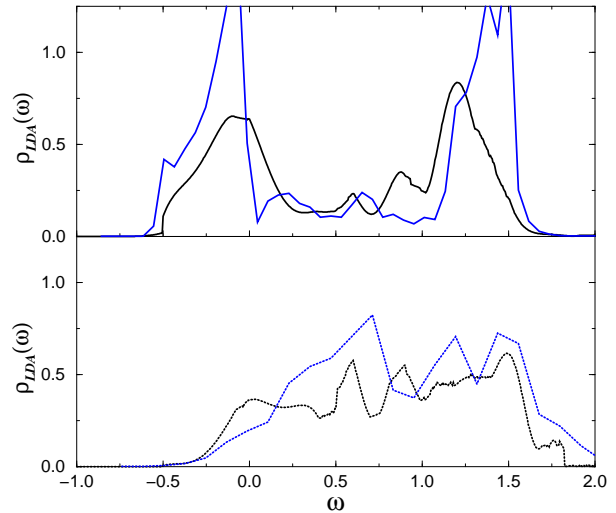}
\caption{
LDA partial density of states for the 
$d_{xy}$ (top) and $d_{yz \pm zx}$ (bottom) orbitals, obtained from 
Ref.~\protect\onlinecite{[17]} (black) and from 
Ref.~\protect\onlinecite{[AG]} (blue). The latter obtained after 
downfolding. }
\label{fig5}
\end{figure}

Finally, another source of discrepancy is the use of different LDA DOS: we
have used the $t_{2g}$ DOS as derived by Korotin {\it et al.}~\cite{[17]}
while the LDA+DMFT(QMC) works use downfolded LDA DOS. As is clear from a 
direct comparison between the two LDA results (see Fig.~\ref{fig5}), 
downfolding the LDA DOS results in an appreciably larger bonding-antibonding 
feature in the $d_{xy}$ DOS. Clearly, the details of the many-body DOS 
depend sensitively on using/not using this device, and the difference 
between our result and those in other works is partly attributable to 
this difference in the corresponding LDA DOS.   
            
While our LDA+DMFT results provide an excellent quantitative description of 
the metallic phase, they give only a very limited description of the $c$-axis 
dimerized insulating state (via structural aspects encoded in the LDA DOS).  
This shortcoming has to do with the intrinsic inability of DMFT itself to 
treat dynamical correlations associated with the quasi-one dimensional 
spin-Peierls transition seen in the I-phase of ${\rm VO_{2}}$.  Thus, a 
quantitative description of the I-phase requires an extension of our 
approach to include the above effects via cluster-DMFT. Such a program is 
under way. 

We can however make a few qualitative remarks on the dimerized phase. 
In our analysis, the $I$ phase is driven by strong 
correlations in concert with 
$\Delta_{\alpha\beta}$. Given this, the existence of long- and short 
V-V distances along $c$ with concomitant $c$-axis spin dimerization would 
follow as a consequence~\cite{[10]} of the $I$ phase and would {\it not} be 
responsible for driving the MIT (incorporation of $\Delta_{\alpha\beta}$
into LDA does not yield the $I$ phase). An analysis along the lines of 
Ref.~\onlinecite{[10]} is possible, but requires more work within our 
formulation. The resulting $c$-axis spin dimerization in the $I$ phase 
should in principle explain the magnetic properties in the $I$ phase. 
This would also have the generic effect of increasing the gap ($E_{G}$) 
to values closer to the experimental value of $0.6$~eV.  More work is 
needed to make quantitative comparison, and this is left for the future.

\section{Conclusion}

In conclusion, in the light of the theoretical constraints imposed by various 
experimental results, we have studied the $T$-driven MIT accompanied by 
abrupt changes in orbital occupations using the ab-initio LDA+DMFT scheme.  
In complete accord with Mott's ideas,~\cite{[9]} the MIT is shown to be of 
the Mott-Hubbard type, driven by an increase in the itinerant carrier density.
The orbital occupations in the PM phase are found to be in excellent 
semiquantitative agreement with indications from recent polarized XAS 
studies. Furthermore, good quantitative agreement with thermodynamic data 
and the one-electron spectrum in {\it both} the $I$ and $M$ phases is 
obtained in the same picture, providing a unified description of the physics 
of ${\rm VO_{2}}$ across the MIT. Such a description can have a wider 
application to other systems~\cite{[6],V2O3} where strong multi-orbital 
correlations in concert with structural details drive a first-order MIT 
under external perturbations. 

\vspace*{-0.5cm}
              
\acknowledgements
We thank L. H. Tjeng for discussions. Work done under the auspices 
of the Sonderforschungsbereich 608 of the Deutsche Forschungsgemeinschaft.
MSL acknowledges financial support from the EPSRC (UK).

\end{document}